# Dissolved $CO_2$ stabilizes dissolution front and increases breakthrough porosity of natural porous materials

Y. Yang*[1], S. Bruns[1], S. L. S. Stipp[1] and H. O. Sørensen[1]

[1] Nano-Science Center, Department of Chemistry, University of Copenhagen, Universitetsparken 5, DK-2100 Copenhagen, Denmark

* yiyang@nano.ku.dk

ABSTRACT

When reactive fluids flow through a dissolving porous medium, conductive channels form, leading to fluid breakthrough. This phenomenon is important in geologic carbon storage, where the dissolution of $CO_2$ in water increases the acidity and produce microstructures significantly different from those in an intact reservoir. We demonstrate the controlling mechanism for the dissolution patterns in natural porous materials. This was done using numerical simulations based on high resolution digital models of North Sea chalk. We tested three model scenarios, and found that aqueous $CO_2$ dissolve porous media homogeneously, leading to large breakthrough porosity. In contrast, $CO_2$–free solution develops elongated convective channels in porous media, known as wormholes, and resulting in small breakthrough porosity. We further show that a homogeneous dissolution pattern appears because the sample size is smaller than the theoretical size of a developing wormhole. The result indicates that the presence of dissolved $CO_2$ expands the reactive subvolume of a porous medium, and thus enhances the geochemical alteration of reservoir structures and might undermine the sealing integrity of caprocks when minerals dissolve.

**Introduction**

A great challenge in predicting the consequences of geologic carbon storage (GCS) is to quantify the interaction between flowing reactive fluids and geologic formations.[1-8] This challenge requires considering a GCS reservoir as a constantly evolving confinement of fluids subject to geochemical and geomechanical instabilities.[9, 10] Reactive infiltration instability (RII) is the morphological instability of a migrating dissolution front to heterogeneities in petrophysical properties. It controls the microstructural evolution of rock in an imposed flow field and is key to the self-organization of natural porous materials.[11-17] Identifying the trigger of this instability, as well as the chemical reactions important for the morphological development,[18, 19] is especially useful for predicting the evolution of sealing integrity of caprocks and the geomechanical deformation of reservoir structures as host rocks are eroded away.

Morphological evolution of porous media caused by reactive infiltration instability can be described qualitative using dissolution patterns (homogeneous, ramified, channelized etc.,[20-22] see also Figure 1), or quantitatively using breakthrough porosity, $\varphi_c$, the macroscopic porosity of a sample when fluid breakthrough occurs.[20, 23, 24] The strength of coupling between mineral dissolution rate and rock permeability is essential to this dynamic process and can be decomposed into 3 types of sensitivity.[21, 25] The first is the sensitivity of flow field to rock microstructure determines how the reactants and products of water-rock interactions are conveyed in a flow field so the chemical reactions are kept away from equilibrium.[26-29] This sensitivity can be affected by, among others, the geometry and texture of pore structures,[30-34] the property and interactions between flowing fluids,[35-38] as well as the body forces exerted by, e.g., gravitational or electrostatic fields.[39] The second sensitivity concerns the kinetic dependence of water-rock interactions to fluid composition, which determines the spatial variations of chemical conversion (i.e., the extent of chemical reaction measured by the relative amount of reactant turned into product).[40] This sensitivity is often complicated by the intrinsic mineral heterogeneity in natural porous media[41-44] and by the large number of aqueous species involved in the reactions.[45, 46] Lastly, the sensitivity of the rock microstructure to chemical conversion closes the loop of the positive feedback leading to RII.[47] This sensitivity reflects the density, molar weight and the solubility of the dissolving or precipitating minerals. If any of the three sensitivities becomes zero, the infiltration instability vanishes and the self-organization of a porous structure ceases.[13, 48, 49]

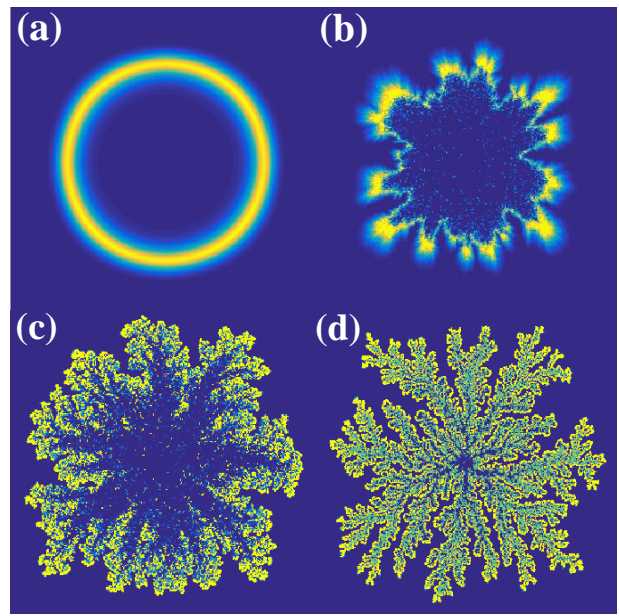

**Figure 1.** A qualitative demonstration of the dissolution front instability. The instability determines the morphology of dissolving porous media. (a) Stable dissolution front: mineral dissolution always takes place at the same distance from the injection point. (b) – (d): increasingly unstable dissolution fronts. The distance between dissolution front and the injection well can vary considerably. Yellow indicates the dissolving regions during the injection of a reactive fluid from the center of the 2D domain. The same amount of reactive fluid is being injected, and snapshots are taken after the same injection time. The instability caused the changes, and a greater instability leads to a higher likelihood of material breakthrough with solid channelisation (wormholing).

Introducing $CO_2$ into a natural porous formation changes many variables mentioned above and thus affects all three types of sensitivity. For example, buoyant gaseous or supercritical $CO_2$ might induce Rayleigh-Taylor instability,[50] influencing the spreading of the reactive aqueous phase and the mixing between fluids.[51] Dissolution of $CO_2$ in water changes many aspects of the kinetics of water-rock interactions. The mineral dissolution rates change because of a lowered pH and a stronger contribution from carbonic and bicarbonate species through surface complexation.[52, 53] The apparent order of reaction, i.e., the sensitivity of reaction rate to reactant concentration, decreases because of the buffering effect of dissolved $CO_2$ as a weak acid.[32] The apparent solubility of minerals can also be affected because of the changed ion activity and the buffered pH. The increased solubility is a thermodynamic driving force for the dissolution reactions. Meanwhile, it can also change the secondary mineral phases and as a result the sensitivity of solid architecture to the extent of reaction.[7, 8, 54, 55] Moreover, the impact of $CO_2$ is related to the way gas is introduced into a geologic setting:[56, 57] direct injection,[58] surface mixing[59] or wellbore mixing[4, 60] all yield distinct pressure and solution compositions.

The dynamics of dissolving microstructures and, consequently, the $CO_2$ effect on the breakthrough properties of natural porous materials remain poorly understood for three reasons. First, structural heterogeneities (e.g., spatial variations in porosity and permeability) are perturbations for reaction front migration. These perturbations are amplified by infiltration instability. Thus, in a series of experiments, the initial microstructure must be the identical for meaningful comparison of results.[19, 30, 31] Second, new models that preserve small scale heterogeneities in numerical simulations are essential.[61-63] The advancement of X-ray imaging has allowed researchers to incorporate real microstructures of natural samples into numerical simulations.[64, 65] However, such an operation usually requires binarisation (segmentation) of the image data. This is because of the limited applicability of governing equations derived from first principles (e.g., Navier-Stokes equations cannot be applied directly to a mixture of solid and fluids).[66, 67] This "hard segmentation" reduces the information contained in each pixel and might erase the important heterogeneities that trigger the unstable migration of the dissolution front.[68, 69] This caveat is especially significant when the imaging resolution is not sufficient to fully resolve the very fine grains and pores in materials such as chalk. A third reason is that simulating microstructural evolution is a free boundary problem of partial differential equations and often requires prohibitive computation even with binarised geometry.[14, 48]

In this study we focus on the effect of dissolved $CO_2$ (as aqueous species) on the microstructural evolution of natural porous materials. The compositions of 3 reactive fluids, corresponding to different model scenarios, were used to flow through the simulation domain. We used high resolution X-ray tomography (25, 50 and 100 nm voxel size) to characterize the microstructures of natural samples from a North Sea chalk formation. Chalk is chosen as a fast dissolving model rock because it is the dominant bedrock in northwest Europe providing oil and gas reservoirs considered for GCS. After computing voxel specific porosities from the 32 bit images, the greyscale datasets were implemented directly into numerical simulations. The simulations allowed us to disentangle the effects of fluid composition from those of different initial microstructures. The simulated spatial and temporal evolutions of rock properties (porosity, permeability and surface area) helped outline the interplay between breakthrough porosity, wormholing and aqueous $CO_2$.

**Methods and Materials**

Three chalk samples from the Hod formation excavated from different locations of North Sea Basin were imaged with X-ray holotomography at 29.49 keV at ID22 of European Synchrotron Research Facility (ESRF).[61] The 3D microstructure were reconstructed from 1999 radiographs at 25, 50 and 100 nm voxel resolutions and processed as described in detail in Bruns et al. Post reconstruction image processing included ring artefact removal according to Jha et al.,[70] followed by iterative nonlocal means denoising[66] and sharpening using the image deconvolution of Wang et al.[71] Greyscale intensities of the voxels were then converted to localized porosity values using linear interpolation between the void phase and the carbonate phase as identified by a fit of a Gaussian mixture model.[67] The obtained greyscale data were imported into numerical simulations based on a previously developed reactor network model.[25] The model describes each imported voxel as a combination of 7 ideal reactors. The volume, permeability and specific surface area of each reactor were assigned according to the voxel size and porosity of neighboring voxels. Each simulation domain consisted of $100^3$ 32-bit voxels, corresponding to 15.625 $\mu m^3$, 125.0 $\mu m^3$ and 1000.0 $\mu m^3$ for the three resolutions. A constant fluid velocity (50 $\mu m/s$) was imposed at the fluid inlet and outlet. Speciation calculations for the three model scenarios (ambient, premixing and direct injection, see Table 1) were conducted using PHREEQC (Version 3) with the *llnl* database.[72, 73] The composition of seawater in scenario II was based on Nordstrom et al.[74] The Peng-Robinson equation of state was used to calculate fugacity coefficients.[75] The B-dot equation was used to calculate the activity coefficients of aqueous species in scenarios II and III.[76] We chose to use the rate law for calcite dissolution from Pokrovsky[52, 53] to approximate the dissolution kinetics of chalk. The surface speciation is calculated based on the aqueous speciation in a closed free-drifting compartment where the aqueous Ca concentration is used as the master variable that determines both the pH and the saturation index of calcite. The rates are then computed and compiled in Figure S1.

**Table 1.** The model scenarios.[74]

| Scenario | ChemID | Temperature (°C) | $P_{CO2}$ (Bar) | Background Electrolytes | Initial pH[*] |
|---|---|---|---|---|---|
| I | 01 | 25 | 1 | n.a. | 3.91 |
| I | 02 | 25 | 0 | n.a. | 3.91 |
| II | 03 | 25 | 25 | Seawater[**] | 3.65 |
| II | 04 | 25 | 0 | Seawater[**] | 3.65 |
| III | 05 | 100 | 250 | 1.0 M NaCl | 3.09 |
| III | 06 | 100 | 0 | 1.0 M NaCl | 3.09 |

[*] Adjusted by hydrochloric acid in the absence of $CO_2$
[**] Nordstrom and others (1979)

**Results and Discussion**

We use the rate of entropy production from fluid friction over the entire sample, $S_P$ (nJ K$^{-1}$ s$^{-1}$), to identify a breakthrough event. $S_P$ quantifies the pressure drop in 3D and is calculated as

$$S_P = -\iiint_V (\mathbf{Q} \cdot \nabla P)\, dV \Big/ T \qquad (1)$$

where $V$ indicates an integration over the entire sample, $\mathbf{Q}$ represents the volumetric flow rate (m$^3$ s$^{-1}$), $P$, the pressure (Pa) and $T$, the temperature (K). When porous media are eroded by reactive fluids, the evolution of $S_P$ depends on the initial microstructure, the kinetics of the dissolution reaction(s) and the flow rate. Figure 2 shows two patterns of $S_P$ evolution: the wormholing pattern, where solid channelisation can be identified, and the homogeneous pattern, where the whole domain dissolves evenly. During wormhole growth, $S_P$ demonstrates one or more rapid drops, signifying the "necking" of regions with high porosities, *i.e.*, small breakthrough of fluid. Necking events are represented by local minima of $S_P$', the first derivative of $S_P$ with respect to time, and can thus be identified by inflection points (Figure 2a). Because of the structural heterogeneity of natural porous media, the critical porosity at which necking occurs, as well as the number of its occurrences, differs for each sample even with the same solution composition. In this study, we report the porosity at the global minimum of $S_P$' as the breakthrough porosity ($\varphi_c$, green arrowed in 2a). This choice does not imply the uniqueness of the necking event during percolation. Figure 2d shows cross sections of a wormholing sample before and after breakthrough and the corresponding spatial patterns of entropy generation. Formation of flow paths with higher permeability lead to the bypassing of fluid from the less porous regions. After breakthrough, a significant pressure drop occurs only at the neck of pores (e.g., $\varphi_a = 0.50$).

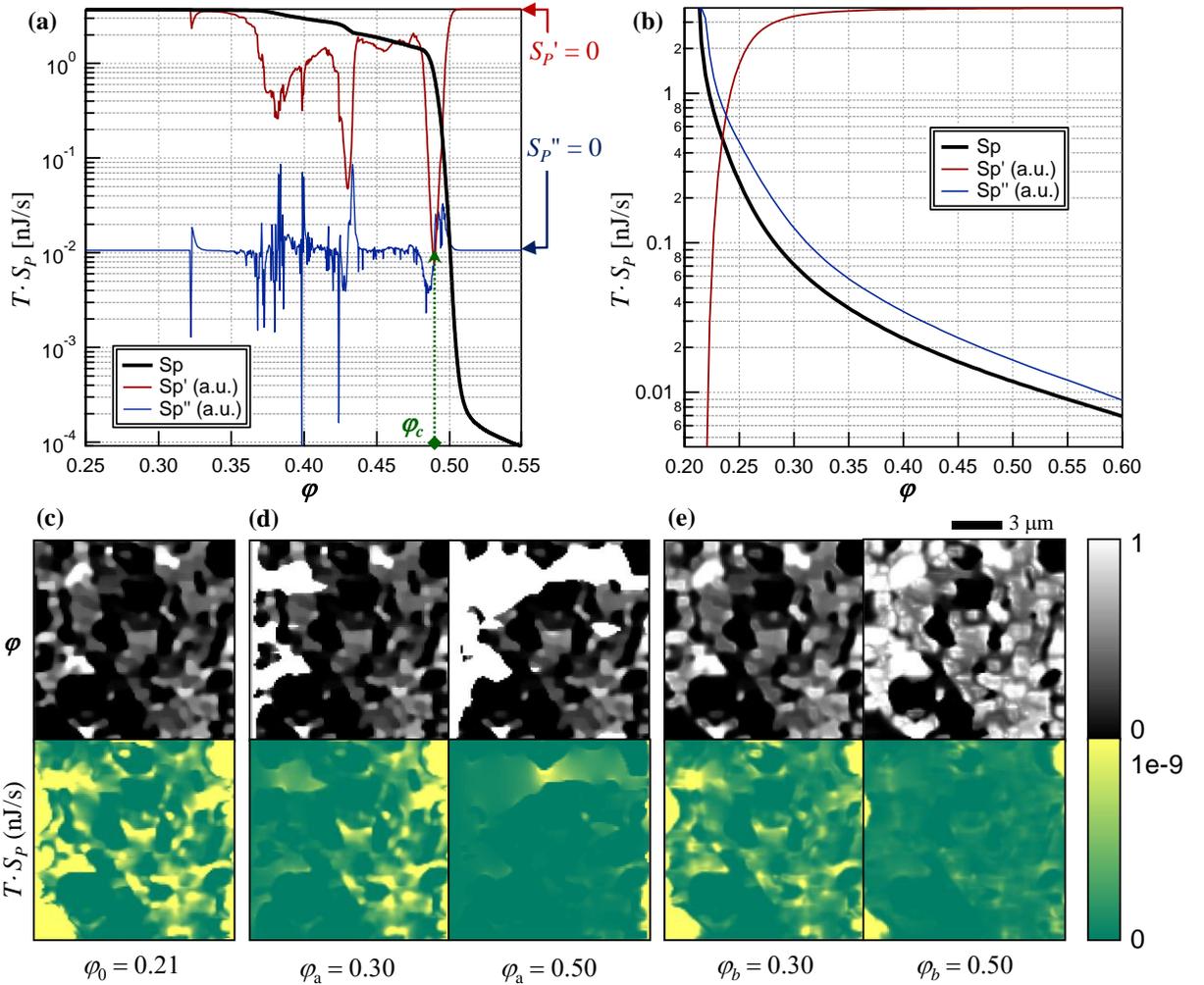

**Figure 2.** Typical temporal and spatial patterns of entropy generation ($S_P$) from fluid friction in developing porous structures. Evolution of $S_P$ in isothermal (a) wormholing and (b) homogeneously dissolving systems. $S_P'$ and $S_P''$ are first and second derivatives of $S_P$ with respect to percolation time. We use the inflection point at the global minimum of $S_P'$ to signify the occurrence of breakthrough (e.g., $\varphi_c$ in a). When $\varphi_c$ is greater than 1, no inflection exists and the medium dissolves homogeneously. (c) – (e) Cross sections of porosity ($\varphi$) and entropy production rate ($T \cdot S_P$) from 3D simulations based on GeoID 1832 (Table S1). Fluid flows from left to right. (c) The initial geometry. (d) A wormholing system before ($\varphi_a = 0.30$) and after ($\varphi_a = 0.50$) breakthrough. (e) A homogeneous dissolution system at the same reaction progresses ($\varphi_b = 0.30$ and $\varphi_b = 0.50$).

Figure 2b shows a temporal pattern of $S_P$ where no abrupt pressure drop can be observed before the depletion of solid material. This pattern indicates a homogeneous dissolution where $S_P'$ increases monotonically. The spatial patterns in Figure 2e reveal that the morphology change does not show strong dependence on the flow direction, i.e. the entire sample behaves as a homogeneous medium despite its geometric complexity, and in different regions porosities decrease at similar rates. Because of the absence of a favored flow path, no significant fluid focusing occurs.

For each of the 40 samples (15 samples each from the 25 nm/pixel and 100 nm/pixel tomogram and 10 samples for the 50 nm/pixel tomogram) we developed 3 model scenarios and simulated the microstructural evolution with and without dissolved $CO_2$, leading to 6 conditions (Table 1) and 240 instances of simulation (Table S1). Scenario I represents the ambient conditions where MilliQ water, equilibrated with 1 bar $CO_2$, is percolating through the porous material at room temperature. In scenario II concentrated $CO_2$ is premixed with seawater at ambient temperature before injected into a geologic formation at intermediate partial pressure of $CO_2$. This operation has been used in, e.g., the CarbFix project in Hellisheidi, Iceland.[60] Scenario III investigates a direct injection of $CO_2$ into deep formations where the supercritical $CO_2$ mixes with brine under reservoir conditions and then migrate away from the injecting well. The results are compared with percolations free of $CO_2$. Hydrochloric acid is added in the absence of $CO_2$ so that solution pairs in each scenario have the same initial pH and therefore a initially comparable calcite dissolution rate. In addition to the breakthrough porosity, we also report the number of pore volumes until breakthrough (# of PV, i.e., the total volume of percolated fluid divided by the initial pore volume of a sample) whenever an $S_P$ inflection can be identified. If a sample dissolves homogeneously, the breakthrough porosity is reported as the difference between its initial porosity and 1. No # of PV is reported for these cases.

Figure 3 shows results of 240 simulations. In 74.2% of all the simulated percolations, dissolved $CO_2$ led to greater breakthrough porosity. This percentage exhibits scenario dependence, being 82.5% for the ambient scenario (I), 77.5% for the premixing scenario (II) and 62.5% for the direct inject scenario (III). In addition, all systems with $CO_2$ need less fluid to breakthrough than the $CO_2$ free systems. The differences are between one to two orders of magnitude. The results also show a weak initial porosity ($\varphi_0$) dependence. The "homogeneous boundaries" (dashed lines) decrease linearly with $\varphi_0$ because they reflect only the total amount of solid initially present and are results of mass balancing. Thus, a completely homogeneous dissolution does not reflect the initial geometric complexity. Also, the number of pore volumes (# of PV) decreases with increasing initial porosity because less fluid needed to breakthrough a more porous material. No significant resolution dependence of the results is observed.

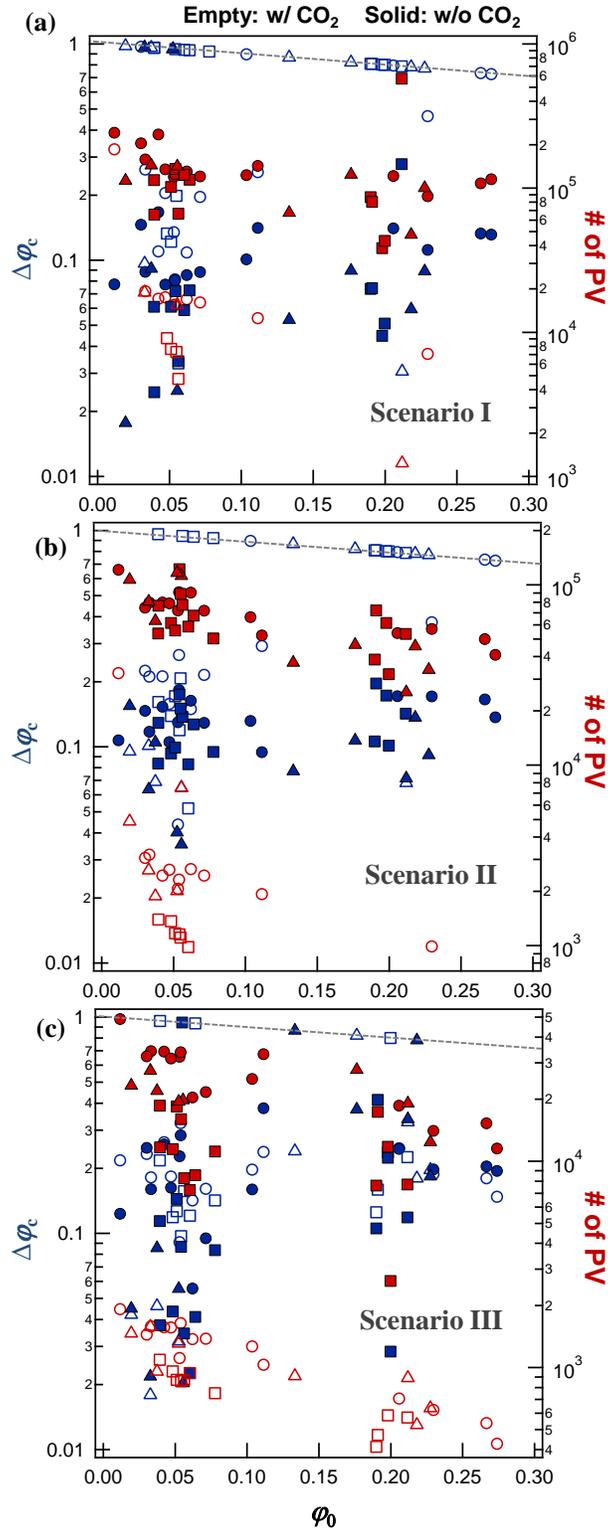

**Figure 3.** Breakthrough porosity and the corresponding number of pore volumes of fluid (# of PV, red symbols). $\Delta\varphi_c$ (blue symbols) represents the difference between the breakthrough porosity ($\varphi_c$) and the initial porosity ($\varphi_0$) of a sample. All samples are cubic and consist of 1

million voxels. The empty symbols show results with $CO_2$ and the solid ones show systems free of $CO_2$. The shape of the symbols represents the voxel size: square – 100 nm, triangle – 50 nm and circle – 25 nm. The gray dashed line near the top of each figure draws the "homogeneous boundary" – corresponding to cases in which no inflection of $S_P$ can be found before solid depletion (*i.e.,* the sample dissolves homogeneously throughout the percolation). For points on this boundary, the number of PV are not shown because of the difficulty in defining "breakthrough". (a) Ambient conditions (Scenario I). (b) Premixing conditions (Scenario II). (c) Direct injection conditions (Scenario III).

Both the differences in the breakthrough porosity ($\Delta\varphi_c$) and the number of pore volumes (# of PV) can be explained by the observed dissolution patterns. Given the same initial structure, the dissolution pattern is determined by distribution of reactants in the medium. Figure 4 shows the decrease of fluid reactivity (measured by reaction rate and pH) as a function of cumulative surface (CS), $\int_0^{\tau_{res}} SSA \cdot dt$, in the 3 model scenarios. The $\tau_{res}$ represents the residence time of fluid in the porous medium (s), and SSA, the specific surface area ($m^{-1}$). The physical significance of CS is the overall surface area the fluid "sees" as it travels through a medium. The curves therefore represent the chemical history of an isolated "fluid parcel", an imaginary constituent of the flowing fluid, travelling along a streamline. Here we assume that the parcel only interacts with the solid on the streamline and does not exchange mass with other fluid parcels. As solid dissolves, the pH and the saturation index (SI) in the fluid parcel increase. Both effects slow down the reaction. The reaction front of the streamline is the position at which the reactivity of the fluid parcel drops to a very small value (e.g., $10^{-6}$ mol/m$^2$/s in Figure 4). This definition suggests the equivalence of residence time and surface area in determining the reaction front. Although specific surface area (SSA) depends exclusively on the sample microstructure, the residence time ($\tau_{res}$) relies on both the microstructure and the fluid flow rate. Consequently, different dissolution patterns can coexist in a system. This is because the patterns reflect the competition between the reaction front and the cumulative surface within a region of interest (ROI). Given the same solution chemistry, one observes a homogeneous dissolution pattern when the fluid leaves the ROI without depleting reactivity. This is typical for fluids with a high apparent mineral solubility. The reaction front of these fluids appears only when residence time is sufficiently long or the specific surface area is large. If, in contrast, the reaction front is reached well before the fluid leaves the ROI, wormholes appear. The residence time can be increased by using a lower flow rate or a bigger sample (i.e., a large ROI). For example, in Figure S2 we show that with an ROI 5 times longer in the flow direction, changing flow rates alone can change the observed dissolution pattern.

Given sample size (*i.e.*, ROI) and flow rate, the CS is fixed, and distinct dissolution patterns are most likely to occur when the ROI contain only one of the two reaction fronts. This is demonstrated by the shaded areas in Figure 4. These areas are left-bounded by the dissolution front of systems without $CO_2$ and right-bounded by those with dissolved $CO_2$. If the CS of a sample falls to the left of the shaded areas, the sample dissolves homogeneously. In contrast, if the CS falls to the right of the shaded areas, wormhole appears regardless of whether $CO_2$ is present. The fluid velocity in this study (50 μm/s) produced distributions of CS that overlap with these shaded areas. Therefore, the widths of the shaded regimes explain the percentages of different patterns observed in the model scenarios. In the ambient scenario (I) the reaction fronts with and without $CO_2$ are farthest from each other (the shaded area is widest in Figure 4a). Without $CO_2$, the ROI is bigger than the reactive volume and the wormholing pattern appears. With dissolved $CO_2$, the opposite is true. As a result 33 of the 40 comparisons (82.5%) show greater breakthrough porosity for systems with $CO_2$. In contrast, the two reaction fronts are very close to each other in the direct injection scenario (III, Figure 4c). Only 62.5% of the simulated cases (25/40) showed that dissolved $CO_2$ increased the breakthrough porosity. The distance between the two reaction fronts is intermediate in the premixing scenario (Figure 4b), and so is the percentage (77.5%).

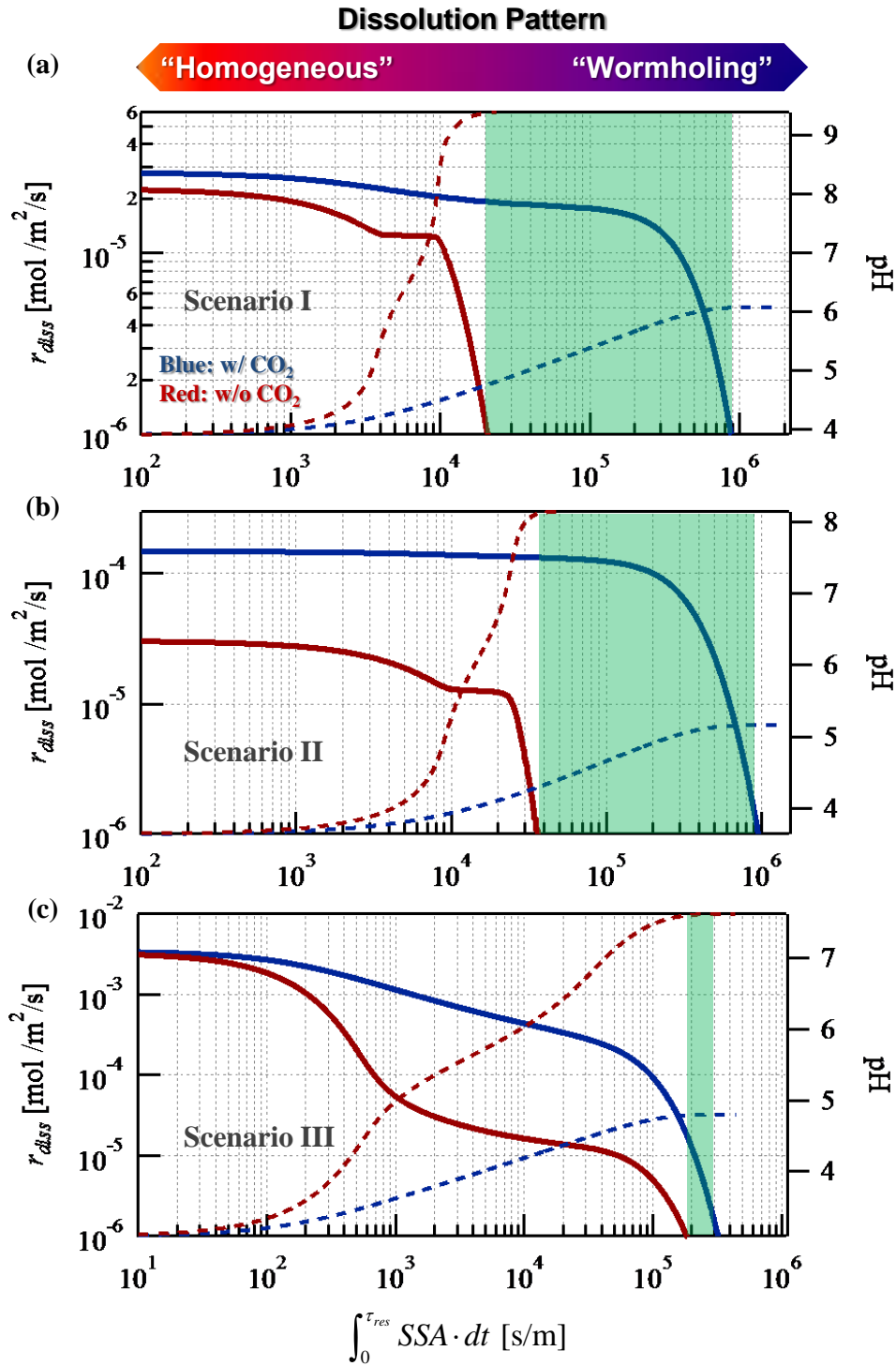

**Figure 4.** Decrease of calcite dissolution rate (solid lines) with cumulative surface area ($\int_0^{\tau_{res}} SSA \cdot dt$) in a free drifting system. The integral represents the overall surface area an isolated "parcel" of reactive fluid experiences before leaving the sample. A greater cumulative surface is more likely to generate wormholing patterns (and vice versa). Shaded regimes are bounded by the dissolution fronts of the two chemical systems. Hence, percolations with and without $CO_2$ in these regimes often produce distinct dissolution patterns. Also shown are the evolutions of pH values (dashed lines). (a) Ambient conditions (Scenario I). (b) Premixing conditions (Scenario II). (c) Direct injection conditions (Scenario III).

Fluid focusing is an important reason why a wormholing pattern relates to lower breakthrough porosity because. Similarly, more effective use of geometric surface for reaction explains why systems with dissolved $CO_2$ need less fluid to breakthrough. Both phenomena are demonstrated in Figure 5, where a case study of microstructural evolution in scenario I is presented. The initial geometry for both simulations is given in Figure S3. The two systems start with the same percolative entropy production rate owing to identical initial geometry. The isothermal $S_P$ for the system with $CO_2$ decreases gradually, indicating a homogeneous dissolution pattern. In contrast, the system without $CO_2$ showed sharp decrease in $S_P$ near $\varphi = 0.4$, suggesting significant necking of pores typical in wormhole growth. $S_P$ without $CO_2$ drops below the one with $CO_2$ after an overall porosity of 0.45, suggesting the bypassing of flow through a fully developed wormhole. The pressure drop after this point is determined by the shape of the channel rather than the porosity of the sample. This redistribution of flow is fluid focusing and is a typical in wormhole growth. Also shown in Figure 5a is the evolution of reactive surface area (RSA). Although the two systems start with exactly the same geometric surface area, their RSA differ by almost one order of magnitude because of the rapid depletion of fluid reactivity in the $CO_2$-free system. The initial increase in RSA is characteristic to a system with infiltration instability. The decrease of RSA is caused by the depletion of solid material. Although RSA in both cases evolve with similar trends, their absolute difference increase with time, contributing to the difference in # of PV at breakthrough.

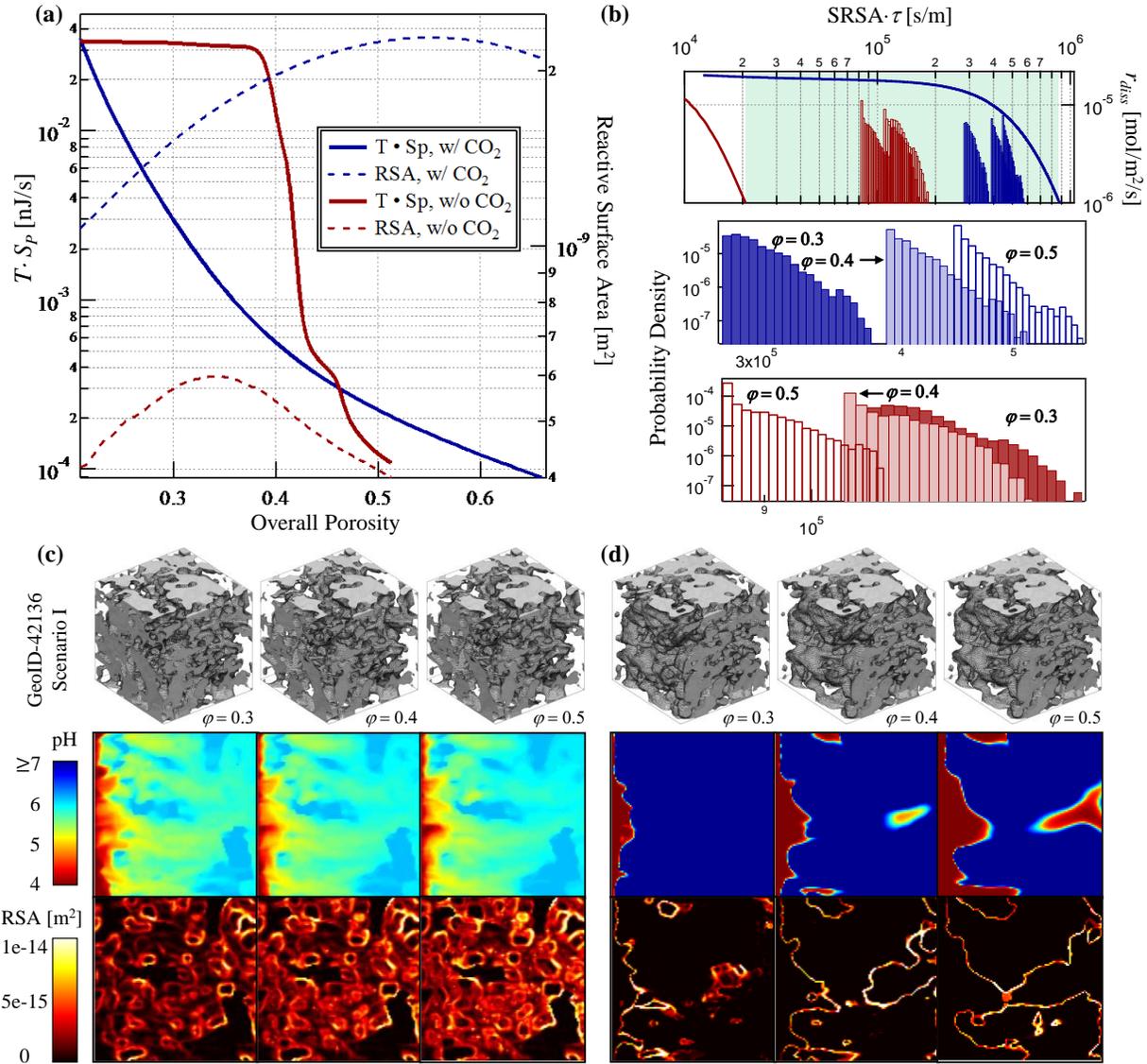

**Figure 5.** A case study of $CO_2$ effect on microstructural evolution (simulations uid-3813 vs. uid-96443). (a) Evolution of isothermal entropy generation rate ($T \cdot S_P$) and reactive surface area (RSA) with overall porosity as an indicator of reaction progress. (b) Distribution of residence time multiplied by specific reactive surface area (SRSA) in the context of calcite dissolution rates in Scenario I. The same shaded regime in Figure 4a is shown. The blue bars show the density function of fluid with $CO_2$ and the red bars, without. Also shown are microstructures, cross sections (10 × 10 μm$^2$) of pH and of reactive surface distribution for percolations (c) with and (d) without $CO_2$.

Figures 5b to 5d compare the evolution of microstructures in the presence and absence of $CO_2$. In 5b the CS of the complete flow field is approximated by the product of specific reactive surface area (SRSA, $m^2/m^3$) and residence time distribution (RTD, giving the probability density of fluid parcels with residence time $\tau$). The same shaded regime in Figure 4a is shown. Both distributions fall entirely into the regime between the two reaction fronts. As a result, the dissolution front with $CO_2$ cannot be observed in the simulation domain because it is beyond the sample size (blue bars). Meanwhile, the dissolution front without $CO_2$ is fully contained in the domain. This is because the sample provides approximately 10 times more cumulative surface required to reach the front (the maximum distance from the injection point to where the dissolution rate drops to zero, the red bars). The development of the distributions at different overall porosities ($\varphi$ = 0.3, 0.4 and 0.5) can be decomposed into two parts. The morphing of their shapes reflects the redistribution of fluid as the micropore evolves. The lower bound of the distribution represents the flow pathways with minimum flow resistance. As the sample dissolves, the mean residence time shifts leftwards indicating that the tendency of fluid to focus on the more permeable pathways. Fluid focusing during wormholing is directly reflected in the standard deviations. With $CO_2$ the standard deviation of the cumulative surface increases from 14769.9 s/m for $\varphi$ = 0.3 to 16137.8 s/m for $\varphi$ = 0.5, suggesting more uniform fluid distribution. In contrast, without $CO_2$ the number decreased from 10071.5 s/m to 6464.07 s/m at the same overall porosities. Therefore, much less geometric surface is in contact with reactive solution during wormhole growth (*e.g.,* spatial distribution of RSA in figures 5c and d). The development of reactive surface determines the horizontal shifting of the distributions. With dissolved $CO_2$, the maximum of RSA appears after an overall porosity of 0.5, and thus increasing the mean of the distribution. In contrast, the overall RSA free of $CO_2$ decreases after an overall porosity of 0.35, shifting the entire distribution of cumulative surface leftward. The spatial distribution of pH shows that the buffering effect of dissolved $CO_2$ increases the apparent solubility of calcite in the fluid and thus the local dissolution rate. Together with greater RSA, the presence of $CO_2$ results in much faster breakthrough (significantly lowered # of PV in all cases) although it increases the breakthrough porosity in 74.2% of the cases.

The $CO_2$ effect on breakthrough porosity has a few implications. Mineral dissolution serves as a trigger for many water-rock interactions. In GCS, dissolution reactions provide cations for carbon mineralization while initiate microstructural evolution that determines the mechanical strength of a formation. Also, changing the solid matrix can the potentially mobilize contaminants. A greater instability of the dissolution front, often associated with wormholing, is favored when geomechanical stability is desirable. This is because lower breakthrough porosity leads to a structure that dissipates injected fluid effectively without removing much solid materials serving as mechanical support. Meanwhile, a wide spread of fluid reactivity often observed in homogeneous dissolution increases the likelihood of contaminant mobilization. Our results show that dissolved $CO_2$ stabilizes the migration of dissolution front by increasing the cumulative surface required for breakthrough, making wormholing less likely within a given sample size. $CO_2$ also shortens dramatically the time for breakthrough. These complications, in addition to brine acidification, may constitute further challenges to engineering GCS.

## ASSOCIATED CONTENT

**Supporting Information**. Calcite dissolution rate based on 3 rate laws available in the literature, for the 3 model scenarios. Cross sections of simulations showing the coexistence of different dissolution patterns within one sample at various flow rates. The initial geometry on which the case study presented in Figure 5 is based. Tabulated results of all simulations.

## AUTHOR INFORMATION

**Corresponding Author**

* yiyang@nano.ku.dk

**Author Contributions**

YY designed the research and conducted the simulations. SB processed the tomography data. HOS and SLSS supervised the research.

## ACKNOWLEDGMENTS

We thank Heikki Suhonen at the ID22 beamline at ESRF (The European Synchrotron) for technical support. We are grateful to F. Engstrøm from Maersk Oil and Gas A/S for providing the sample. YY deeply appreciates Dr. K. Dideriksen for reviewing the manuscript and for the thought provoking discussions. Funding for this project was provided by the Innovation Fund Denmark, through the CINEMA project, the Innovation Fund Denmark and Maersk Oil and Gas A/S, through the $P^3$ project as well as the European Commission, Horizon 2020 Research and Innovation Programme under the Marie Sklodowska-Curie Grant Agreement No 653241. We thank the Danish Council for Independent Research for support for synchrotron beamtime through DANSCATT.

**With CO$_2$**  **Without CO$_2$**

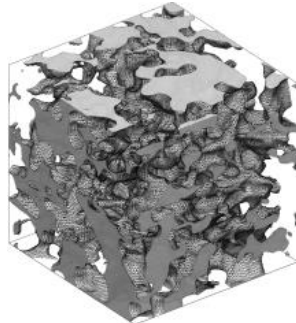 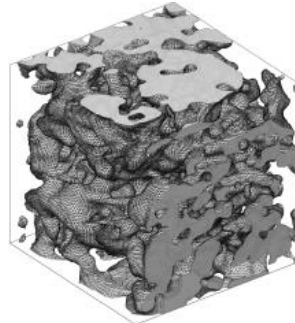

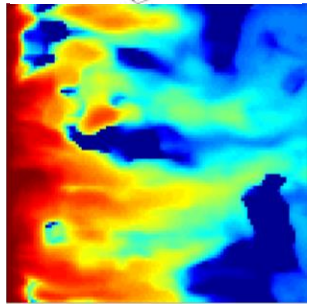 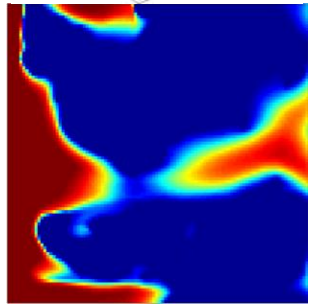

Supporting Information for

Dissolved CO$_2$ stabilizes dissolution front and increases breakthrough porosity of natural porous materials

*Y. Yang\*, S. Bruns, S. L. S. Stipp and H. O. Sørensen*

Nano-Science Center, Department of Chemistry, University of Copenhagen

Universitetsparken 5, DK-2100 Copenhagen, Denmark

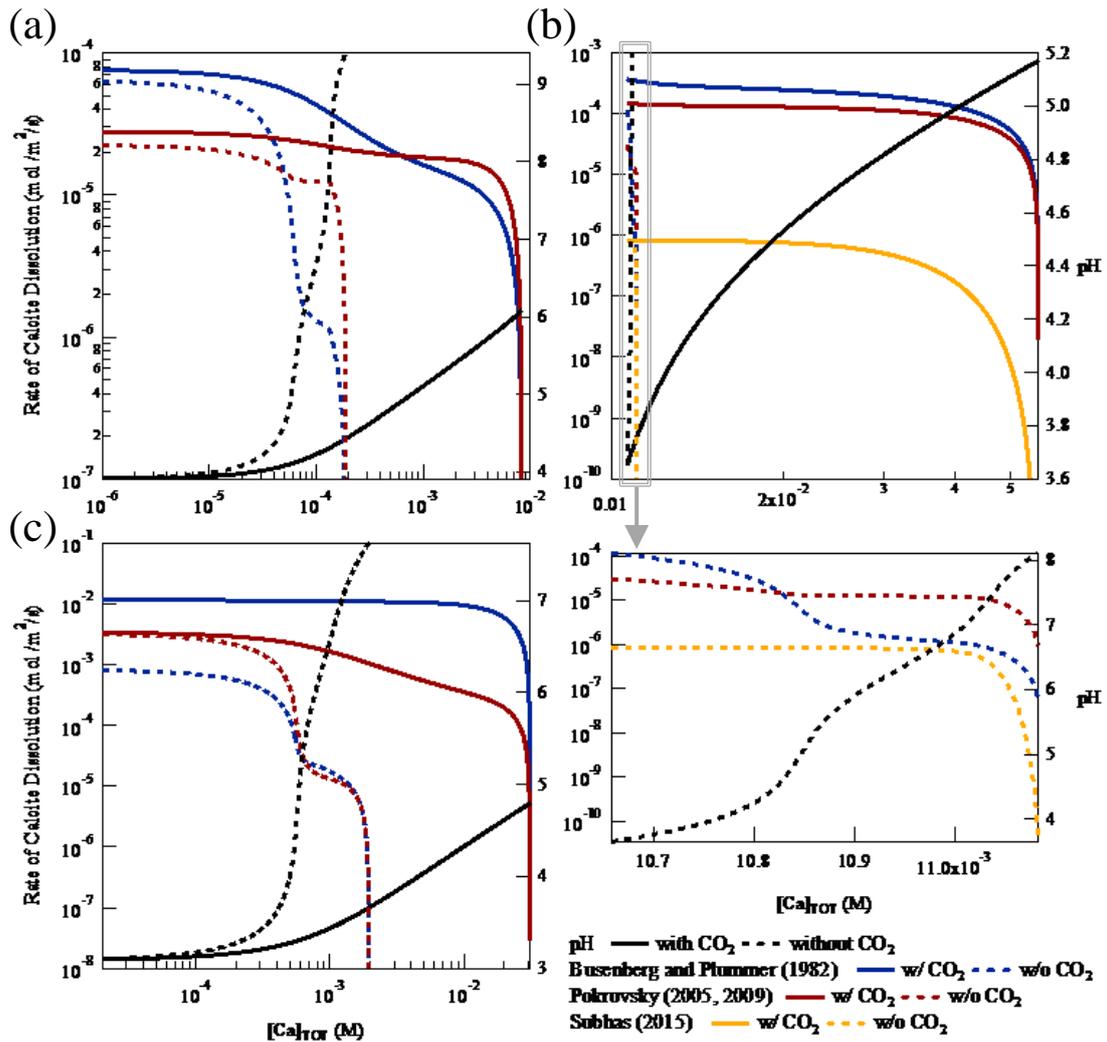

**Figure S1.** Rate of calcite dissolution in a closed (free-drifting) system based on Busenberg and Plummer,[1] Pokrovsky[2, 3] and Subhas[4] in the model scenarios: (a) Ambient, (b) Premixing and (c) Direct injection.

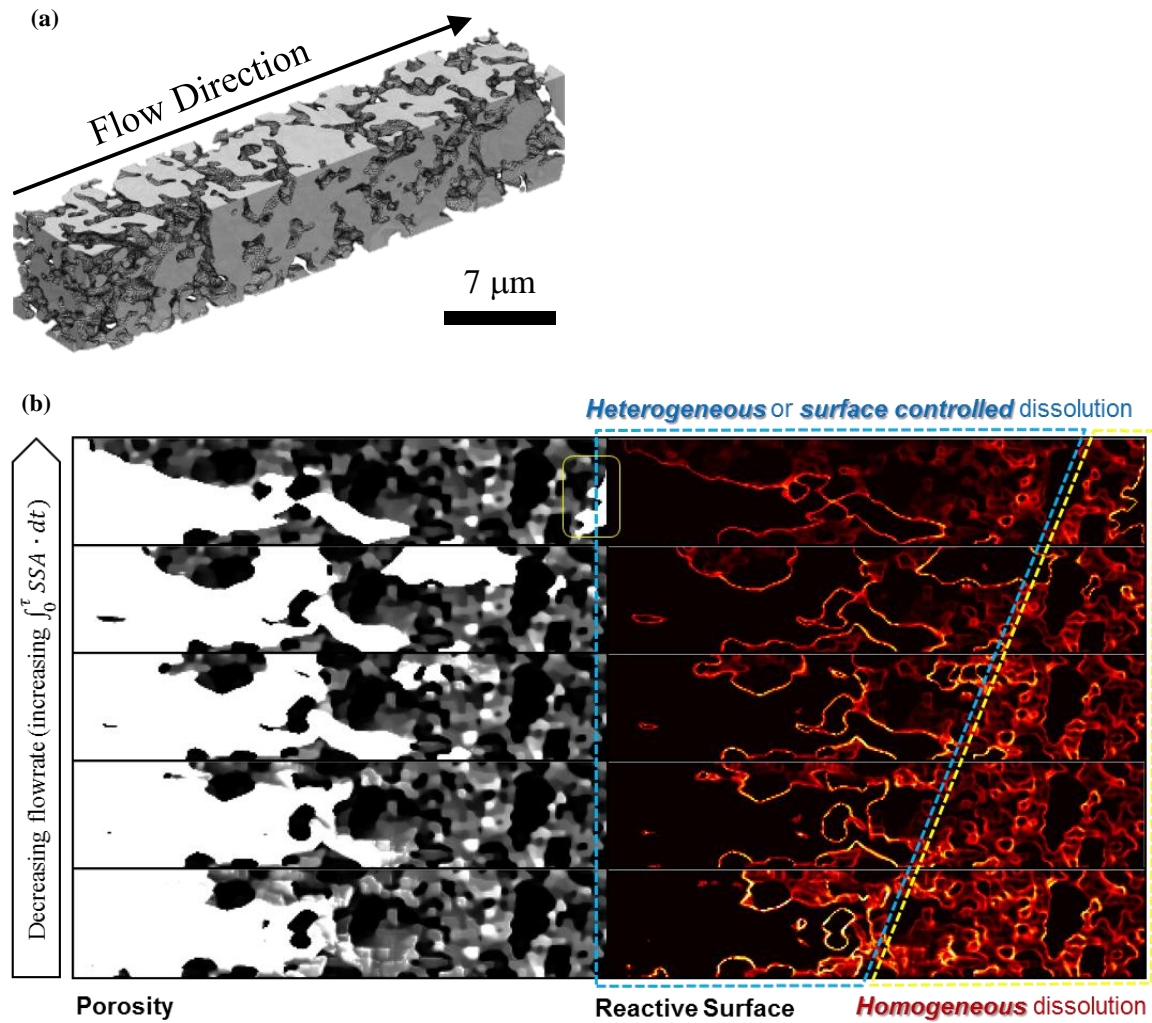

**Figure S2.** Simulations showing the coexistence of a homogeneously dissolving region and a heterogeneous or surface controlled (wormholing) region within a sample. The sample is longer in the flow direction than other samples so the cumulative surface is sufficiently large to contain the reaction front. Decreasing flowrate further increases the residence time and thus the cumulative surface. A first order rate law was used. (a) Simulation setup. The fluid flows from left to right. (b) Microstructures of the same sample at $\varphi = 0.5$ with varying flowrates. Note that with the lowest flowrate (top row in b) the breakthrough has occurred (yellow box).

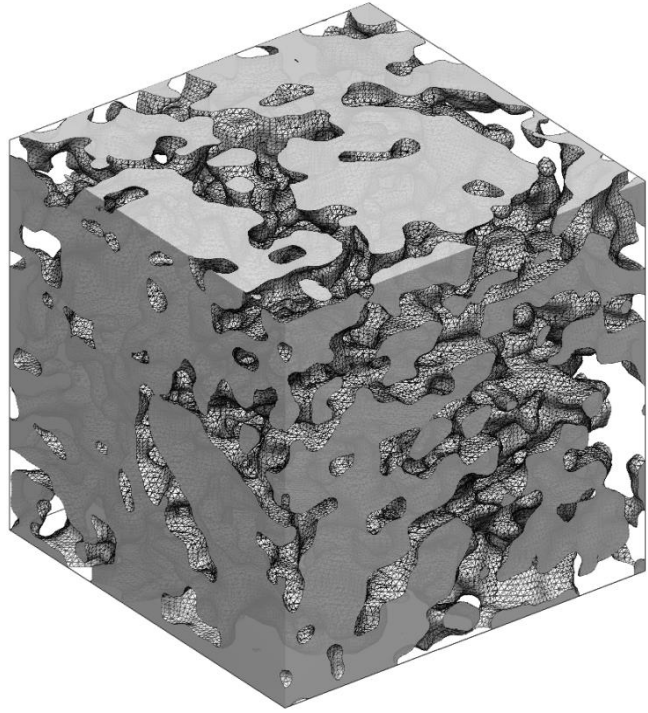

**Figure S3.** Initial geometry of the simulations shown in Figure 5 (voxel size: 100 nm, GeoID-42136, $\varphi_0 = 0.199$). Fluid flows from the lower left face towards the upper right face.

**Table S1.** Results of simulations conducted in this study. GeoID is a unique number assigned to each sample used as an initial microstructure in simulations. $l$ indicates the voxel resolution. $\varphi_0$ represents the initial porosity of the sample. ChemID is a unique number assigned to each chemical scenario (Table 1). uid is a unique number assigned to each simulation. $\varphi_c$ indicates the breakthrough porosity. $PV_c$ is the number of pore volumes of fluid to breakthrough. $\delta\varphi_c$ represents the difference between the initial and the breakthrough porosity. "n.a." indicates a sample dissolves homogeneously.

| GeoID | $l$ (nm) | $\varphi_0$ | ChemID | uid | $\varphi_c$ | $PV_c$ | $\delta\varphi_c$ |
|---|---|---|---|---|---|---|---|
| 9435 | 25 | 0.2665 | 01 | 97658 | n.a. | n.a. | 0.7335 |
| 9435 | 25 | 0.2665 | 02 | 29203 | 0.3994 | 1.08E+05 | 0.1329 |
| 9435 | 25 | 0.2665 | 03 | 24461 | n.a. | n.a. | 0.7335 |
| 9435 | 25 | 0.2665 | 04 | 93567 | 0.4320 | 5.00E+04 | 0.1655 |
| 9435 | 25 | 0.2665 | 05 | 30768 | 0.4465 | 5.38E+02 | 0.1800 |
| 9435 | 25 | 0.2665 | 06 | 84652 | 0.4707 | 1.53E+04 | 0.2042 |
| 52235 | 25 | 0.2057 | 01 | 2736 | n.a. | n.a. | 0.7943 |
| 52235 | 25 | 0.2057 | 02 | 72086 | 0.3460 | 1.21E+05 | 0.1403 |
| 52235 | 25 | 0.2057 | 03 | 78443 | n.a. | n.a. | 0.7943 |
| 52235 | 25 | 0.2057 | 04 | 15382 | 0.3772 | 5.41E+04 | 0.1715 |
| 52235 | 25 | 0.2057 | 05 | 53764 | 0.4529 | 7.07E+02 | 0.2472 |
| 52235 | 25 | 0.2057 | 06 | 880 | 0.4516 | 1.86E+04 | 0.2459 |
| 53419 | 25 | 0.2297 | 01 | 63248 | 0.6931 | 7.07E+03 | 0.4634 |
| 53419 | 25 | 0.2297 | 02 | 75810 | 0.3413 | 8.79E+04 | 0.1116 |
| 53419 | 25 | 0.2297 | 03 | 18981 | 0.6061 | 9.90E+02 | 0.3764 |
| 53419 | 25 | 0.2297 | 04 | 67963 | 0.4006 | 5.69E+04 | 0.1709 |
| 53419 | 25 | 0.2297 | 05 | 53091 | 0.4175 | 6.23E+02 | 0.1878 |
| 53419 | 25 | 0.2297 | 06 | 25978 | 0.4273 | 1.40E+04 | 0.1976 |
| 60917 | 25 | 0.0530 | 01 | 42252 | 0.1878 | 1.54E+04 | 0.1348 |
| 60917 | 25 | 0.0530 | 02 | 43265 | 0.1303 | 1.20E+05 | 0.0773 |
| 60917 | 25 | 0.0530 | 03 | 97390 | 0.0967 | 2.07E+03 | 0.0437 |
| 60917 | 25 | 0.0530 | 04 | 35121 | 0.1829 | 7.22E+04 | 0.1299 |
| 60917 | 25 | 0.0530 | 05 | 12273 | 0.1441 | 1.11E+03 | 0.0911 |
| 60917 | 25 | 0.0530 | 06 | 6078 | 0.2802 | 3.21E+04 | 0.2272 |
| 82235 | 25 | 0.2739 | 01 | 25689 | n.a. | n.a. | 0.7261 |
| 82235 | 25 | 0.2739 | 02 | 65550 | 0.4053 | 1.15E+05 | 0.1314 |
| 82235 | 25 | 0.2739 | 03 | 19540 | n.a. | n.a. | 0.7261 |
| 82235 | 25 | 0.2739 | 04 | 1879 | 0.4108 | 4.09E+04 | 0.1369 |
| 82235 | 25 | 0.2739 | 05 | 55671 | 0.4215 | 4.27E+02 | 0.1476 |
| 82235 | 25 | 0.2739 | 06 | 62117 | 0.4684 | 1.15E+04 | 0.1945 |
| 4888 | 50 | 0.2276 | 01 | 75994 | n.a. | n.a. | 0.7724 |
| 4888 | 50 | 0.2276 | 02 | 93376 | 0.3165 | 1.00E+05 | 0.0889 |
| 4888 | 50 | 0.2276 | 03 | 50855 | n.a. | n.a. | 0.7724 |
| 4888 | 50 | 0.2276 | 04 | 41784 | 0.3192 | 3.38E+04 | 0.0916 |
| 4888 | 50 | 0.2276 | 05 | 73916 | 0.4243 | 6.36E+02 | 0.1967 |
| 4888 | 50 | 0.2276 | 06 | 52242 | 0.4112 | 1.24E+04 | 0.1836 |
| 9087 | 50 | 0.2118 | 01 | 22839 | 0.2424 | 1.24E+03 | 0.0306 |
| 9087 | 50 | 0.2118 | 03 | 90790 | 0.2800 | 3.21E+02 | 0.0682 |
| 9087 | 50 | 0.2118 | 04 | 7930 | 0.2834 | 2.54E+04 | 0.0716 |
| 9087 | 50 | 0.2118 | 05 | 24151 | 0.5389 | 8.91E+02 | 0.3271 |
| 9087 | 50 | 0.2118 | 06 | 22557 | 0.5496 | 1.91E+04 | 0.3378 |
| 30168 | 50 | 0.2182 | 01 | 57121 | n.a. | n.a. | 0.7818 |
| 30168 | 50 | 0.2182 | 02 | 10976 | 0.3628 | 1.72E+05 | 0.1446 |
| 30168 | 50 | 0.2182 | 03 | 74869 | n.a. | n.a. | 0.7818 |
| 30168 | 50 | 0.2182 | 04 | 54238 | 0.3547 | 4.57E+04 | 0.1365 |
| 30168 | 50 | 0.2182 | 05 | 57041 | 0.3982 | 5.27E+02 | 0.1800 |
| 30168 | 50 | 0.2182 | 06 | 5281 | n.a. | n.a. | 0.7818 |
| 88470 | 50 | 0.1333 | 01 | 70913 | n.a. | n.a. | 0.8667 |
| 88470 | 50 | 0.1333 | 02 | 18747 | 0.1864 | 6.76E+04 | 0.0531 |
| 88470 | 50 | 0.1333 | 03 | 73468 | n.a. | n.a. | 0.8667 |
| 88470 | 50 | 0.1333 | 04 | 82604 | 0.2105 | 3.71E+04 | 0.0772 |
| 88470 | 50 | 0.1333 | 05 | 79565 | 0.3733 | 9.11E+02 | 0.2400 |
| 88470 | 50 | 0.1333 | 06 | 20898 | n.a. | n.a. | 0.8667 |
| 98562 | 50 | 0.1763 | 01 | 21682 | n.a. | n.a. | 0.8237 |
| 98562 | 50 | 0.1763 | 02 | 79784 | 0.2658 | 1.24E+05 | 0.0895 |
| 98562 | 50 | 0.1763 | 03 | 8642 | n.a. | n.a. | 0.8237 |
| 98562 | 50 | 0.1763 | 04 | 36574 | 0.2835 | 4.66E+04 | 0.1072 |
| 98562 | 50 | 0.1763 | 05 | 97447 | n.a. | n.a. | 0.8237 |
| 98562 | 50 | 0.1763 | 06 | 91245 | 0.5506 | 2.78E+04 | 0.3743 |
| 1832 | 100 | 0.2116 | 01 | 43103 | n.a. | n.a. | 0.7884 |
| 1832 | 100 | 0.2116 | 02 | 29158 | 0.4900 | 5.75E+05 | 0.2784 |
| 1832 | 100 | 0.2116 | 03 | 96536 | n.a. | n.a. | 0.7884 |
| 1832 | 100 | 0.2116 | 04 | 48411 | 0.3540 | 5.34E+04 | 0.1424 |
| 1832 | 100 | 0.2116 | 05 | 82856 | 0.4370 | 5.72E+02 | 0.2254 |
| 1832 | 100 | 0.2116 | 06 | 37819 | 0.3301 | 7.73E+03 | 0.1185 |
| 36487 | 100 | 0.1899 | 01 | 20530 | n.a. | n.a. | 0.8101 |
| 36487 | 100 | 0.1899 | 02 | 60354 | 0.2636 | 8.65E+04 | 0.0737 |
| 36487 | 100 | 0.1899 | 03 | 29377 | n.a. | n.a. | 0.8101 |
| 36487 | 100 | 0.1899 | 04 | 45197 | 0.2957 | 3.86E+04 | 0.1058 |
| 36487 | 100 | 0.1899 | 05 | 16956 | 0.3151 | 4.13E+02 | 0.1252 |
| 36487 | 100 | 0.1899 | 06 | 73427 | 0.2954 | 7.62E+03 | 0.1055 |
| 42136 | 100 | 0.1979 | 01 | 3813 | n.a. | n.a. | 0.8021 |
| 42136 | 100 | 0.1979 | 02 | 96443 | 0.3936 | 2.79E+05 | 0.1957 |
| 42136 | 100 | 0.1979 | 03 | 23133 | n.a. | n.a. | 0.8021 |
| 42136 | 100 | 0.1979 | 04 | 28767 | 0.3707 | 6.15E+04 | 0.1728 |
| 42136 | 100 | 0.1979 | 05 | 62224 | 0.4345 | 5.87E+02 | 0.2366 |
| 42136 | 100 | 0.1979 | 06 | 16337 | 0.4215 | 1.17E+04 | 0.2236 |
| 60234 | 100 | 0.1997 | 01 | 26582 | n.a. | n.a. | 0.8003 |
| 60234 | 100 | 0.1997 | 02 | 43249 | 0.2506 | 4.31E+04 | 0.0509 |
| 60234 | 100 | 0.1997 | 03 | 62411 | n.a. | n.a. | 0.8003 |
| 60234 | 100 | 0.1997 | 04 | 64051 | 0.3008 | 3.19E+04 | 0.1011 |
| 60234 | 100 | 0.1997 | 05 | 24212 | n.a. | n.a. | 0.8003 |
| 60234 | 100 | 0.1997 | 06 | 94579 | 0.2281 | 2.63E+03 | 0.0284 |
| 82949 | 100 | 0.1909 | 01 | 53545 | n.a. | n.a. | 0.8091 |
| 82949 | 100 | 0.1909 | 02 | 69476 | 0.2651 | 8.04E+04 | 0.0742 |
| 82949 | 100 | 0.1909 | 03 | 77677 | n.a. | n.a. | 0.8091 |
| 82949 | 100 | 0.1909 | 04 | 20012 | 0.3872 | 7.21E+04 | 0.1963 |
| 82949 | 100 | 0.1909 | 05 | 46003 | 0.3501 | 4.71E+02 | 0.1592 |
| 82949 | 100 | 0.1909 | 06 | 91577 | 0.6053 | 1.73E+04 | 0.4144 |
| 29098 | 25 | 0.0331 | 01 | 37273 | 0.2958 | 1.93E+04 | 0.2627 |
| 29098 | 25 | 0.0331 | 02 | 67330 | 0.1214 | 1.58E+05 | 0.0883 |
| 29098 | 25 | 0.0331 | 03 | 68680 | 0.2442 | 3.19E+03 | 0.2111 |
| 29098 | 25 | 0.0331 | 04 | 68724 | 0.1505 | 7.97E+04 | 0.1174 |
| 29098 | 25 | 0.0331 | 05 | 49808 | 0.2148 | 1.58E+03 | 0.1817 |
| 29098 | 25 | 0.0331 | 06 | 52768 | 0.1932 | 3.42E+04 | 0.1601 |
| 34437 | 25 | 0.0538 | 01 | 24177 | 0.1354 | 1.55E+04 | 0.0816 |
| 34437 | 25 | 0.0538 | 02 | 42957 | 0.1346 | 1.25E+05 | 0.0808 |
| 34437 | 25 | 0.0538 | 03 | 52245 | 0.3195 | 2.32E+03 | 0.2657 |
| 34437 | 25 | 0.0538 | 04 | 17494 | 0.2368 | 9.12E+04 | 0.1830 |
| 34437 | 25 | 0.0538 | 05 | 70270 | 0.3776 | 1.64E+03 | 0.3238 |
| 34437 | 25 | 0.0538 | 06 | 40030 | 0.3375 | 3.38E+04 | 0.2837 |
| 38978 | 25 | 0.0117 | 01 | 38486 | 0.0194 | 1.86E+05 | 0.0077 |
| 38978 | 25 | 0.0117 | 02 | 45174 | 0.0890 | 2.43E+05 | 0.0773 |
| 38978 | 25 | 0.0117 | 03 | 18980 | 0.0210 | 3.24E+04 | 0.0093 |
| 38978 | 25 | 0.0117 | 04 | 18540 | 0.1187 | 1.21E+05 | 0.1070 |
| 38978 | 25 | 0.0117 | 05 | 33319 | 0.2294 | 1.92E+03 | 0.2177 |
| 38978 | 25 | 0.0117 | 06 | 42172 | 0.1350 | 4.89E+04 | 0.1233 |
| 49596 | 25 | 0.0620 | 01 | 97591 | 0.1707 | 1.70E+04 | 0.1087 |
| 49596 | 25 | 0.0620 | 02 | 60986 | 0.1473 | 1.30E+05 | 0.0853 |
| 49596 | 25 | 0.0620 | 03 | 7369 | 0.2113 | 2.65E+03 | 0.1493 |
| 49596 | 25 | 0.0620 | 04 | 56418 | 0.2253 | 9.06E+04 | 0.1633 |
| 49596 | 25 | 0.0620 | 05 | 52410 | 0.2040 | 1.37E+03 | 0.1420 |
| 49596 | 25 | 0.0620 | 06 | 75510 | 0.1176 | 2.04E+04 | 0.0556 |
| 60513 | 25 | 0.1035 | 01 | 85735 | n.a. | n.a. | 0.8965 |
| 60513 | 25 | 0.1035 | 02 | 5941 | 0.2044 | 1.23E+05 | 0.1009 |
| 60513 | 25 | 0.1035 | 03 | 48943 | n.a. | n.a. | 0.8965 |
| 60513 | 25 | 0.1035 | 04 | 44765 | 0.2353 | 6.63E+04 | 0.1318 |
| 60513 | 25 | 0.1035 | 05 | 87848 | 0.3005 | 1.26E+03 | 0.1970 |
| 60513 | 25 | 0.1035 | 06 | 77834 | 0.2630 | 2.51E+04 | 0.1595 |
| 49707 | 100 | 0.0640 | 01 | 84367 | n.a. | n.a. | 0.9360 |
| 49707 | 100 | 0.0640 | 02 | 48761 | 0.1365 | 1.14E+05 | 0.0725 |
| 49707 | 100 | 0.0640 | 03 | 77226 | n.a. | n.a. | 0.9360 |
| 49707 | 100 | 0.0640 | 04 | 30262 | 0.1909 | 6.76E+04 | 0.1269 |
| 49707 | 100 | 0.0640 | 05 | 54920 | n.a. | n.a. | 0.9360 |
| 49707 | 100 | 0.0640 | 06 | 930 | 0.1051 | 8.56E+03 | 0.0411 |
| 62954 | 100 | 0.0562 | 01 | 94868 | 0.0895 | 4.75E+03 | 0.0333 |
| 62954 | 100 | 0.0562 | 02 | 76896 | 0.0903 | 6.64E+04 | 0.0341 |
| 62954 | 100 | 0.0562 | 03 | 26211 | n.a. | n.a. | 0.9438 |
| 62954 | 100 | 0.0562 | 04 | 64094 | 0.1936 | 7.75E+04 | 0.1374 |
| 62954 | 100 | 0.0562 | 05 | 10574 | 0.2128 | 8.69E+02 | 0.1566 |
| 62954 | 100 | 0.0562 | 06 | 61450 | 0.0906 | 8.27E+03 | 0.0344 |
| 63250 | 100 | 0.0395 | 01 | 16746 | n.a. | n.a. | 0.9605 |
| 63250 | 100 | 0.0395 | 02 | 39601 | 0.0640 | 6.56E+04 | 0.0245 |
| 63250 | 100 | 0.0395 | 03 | 77205 | 0.2008 | 1.39E+03 | 0.1613 |
| 63250 | 100 | 0.0395 | 04 | 88867 | 0.1684 | 7.66E+04 | 0.1289 |
| 63250 | 100 | 0.0395 | 05 | 98700 | n.a. | n.a. | 0.9605 |
| 63250 | 100 | 0.0395 | 06 | 47834 | 0.0771 | 1.17E+04 | 0.0376 |
| 73318 | 100 | 0.0777 | 01 | 22017 | n.a. | n.a. | 0.9223 |
| 73318 | 100 | 0.0777 | 02 | 27294 | 0.1607 | 1.33E+05 | 0.0830 |
| 73318 | 100 | 0.0777 | 03 | 29243 | n.a. | n.a. | 0.9223 |
| 73318 | 100 | 0.0777 | 04 | 3003 | 0.1723 | 5.05E+04 | 0.0946 |
| 73318 | 100 | 0.0777 | 05 | 20241 | 0.2197 | 7.51E+02 | 0.1420 |
| 73318 | 100 | 0.0777 | 06 | 32205 | 0.1613 | 1.12E+04 | 0.0836 |
| 82835 | 100 | 0.0541 | 01 | 31467 | n.a. | n.a. | 0.9459 |
| 82835 | 100 | 0.0541 | 02 | 3724 | 0.1259 | 1.36E+05 | 0.0718 |
| 82835 | 100 | 0.0541 | 03 | 37369 | 0.1735 | 1.15E+03 | 0.1194 |
| 82835 | 100 | 0.0541 | 04 | 80241 | 0.2285 | 1.22E+05 | 0.1744 |
| 82835 | 100 | 0.0541 | 05 | 49277 | 0.1512 | 8.60E+02 | 0.0971 |
| 82835 | 100 | 0.0541 | 06 | 50508 | 0.1407 | 1.60E+04 | 0.0866 |
| 14228 | 25 | 0.0302 | 01 | 79728 | n.a. | n.a. | 0.9698 |
| 14228 | 25 | 0.0302 | 02 | 9236 | 0.1761 | 2.04E+05 | 0.1459 |
| 14228 | 25 | 0.0302 | 03 | 92737 | 0.2547 | 3.06E+03 | 0.2245 |
| 14228 | 25 | 0.0302 | 04 | 80205 | 0.1767 | 7.47E+04 | 0.1465 |
| 14228 | 25 | 0.0302 | 05 | 33626 | 0.2631 | 1.44E+03 | 0.2329 |
| 14228 | 25 | 0.0302 | 06 | 34807 | 0.2786 | 3.23E+04 | 0.2484 |
| 20374 | 25 | 0.0423 | 01 | 85216 | 0.1522 | 1.71E+04 | 0.1099 |
| 20374 | 25 | 0.0423 | 02 | 783 | 0.2090 | 2.35E+05 | 0.1667 |
| 20374 | 25 | 0.0423 | 03 | 44418 | 0.2538 | 2.44E+03 | 0.2115 |
| 20374 | 25 | 0.0423 | 04 | 24344 | 0.1954 | 7.98E+04 | 0.1531 |
| 20374 | 25 | 0.0423 | 05 | 5146 | 0.3064 | 1.57E+03 | 0.2641 |
| 20374 | 25 | 0.0423 | 06 | 68837 | 0.3000 | 3.40E+04 | 0.2577 |
| 43457 | 25 | 0.0470 | 01 | 74069 | 0.2516 | 1.75E+04 | 0.2046 |
| 43457 | 25 | 0.0470 | 02 | 42311 | 0.1242 | 1.35E+05 | 0.0772 |
| 43457 | 25 | 0.0470 | 03 | 52265 | 0.2050 | 2.63E+03 | 0.1580 |
| 43457 | 25 | 0.0470 | 04 | 42654 | 0.1522 | 7.89E+04 | 0.1052 |
| 43457 | 25 | 0.0470 | 05 | 97901 | 0.2300 | 1.57E+03 | 0.1830 |
| 43457 | 25 | 0.0470 | 06 | 68838 | 0.2097 | 3.15E+04 | 0.1627 |
| 52989 | 25 | 0.1114 | 01 | 78241 | 0.3666 | 1.25E+04 | 0.2552 |
| 52989 | 25 | 0.1114 | 02 | 65558 | 0.2526 | 1.43E+05 | 0.1412 |
| 52989 | 25 | 0.1114 | 03 | 52250 | 0.4042 | 1.93E+03 | 0.2928 |
| 52989 | 25 | 0.1114 | 04 | 44637 | 0.2059 | 5.25E+04 | 0.0945 |
| 52989 | 25 | 0.1114 | 05 | 56626 | 0.3494 | 1.03E+03 | 0.2380 |
| 52989 | 25 | 0.1114 | 06 | 32553 | 0.4901 | 3.31E+04 | 0.3787 |
| 93330 | 25 | 0.0712 | 01 | 97509 | 0.2673 | 1.61E+04 | 0.1961 |
| 93330 | 25 | 0.0712 | 02 | 72293 | 0.1593 | 1.20E+05 | 0.0881 |
| 93330 | 25 | 0.0712 | 03 | 89618 | 0.2865 | 2.44E+03 | 0.2153 |
| 93330 | 25 | 0.0712 | 04 | 23495 | 0.2002 | 7.19E+04 | 0.1290 |
| 93330 | 25 | 0.0712 | 05 | 72341 | 0.2321 | 1.38E+03 | 0.1609 |
| 93330 | 25 | 0.0712 | 06 | 96375 | 0.1661 | 2.17E+04 | 0.0949 |
| 2891 | 50 | 0.0194 | 01 | 4686 | n.a. | n.a. | 0.9806 |
| 2891 | 50 | 0.0194 | 02 | 10882 | 0.0371 | 1.13E+05 | 0.0177 |
| 2891 | 50 | 0.0194 | 03 | 77761 | 0.1148 | 4.88E+03 | 0.0954 |
| 2891 | 50 | 0.0194 | 04 | 51035 | 0.1741 | 1.07E+05 | 0.1547 |
| 2891 | 50 | 0.0194 | 05 | 84515 | 0.0615 | 1.46E+03 | 0.0421 |
| 2891 | 50 | 0.0194 | 06 | 6472 | 0.0643 | 2.33E+04 | 0.0449 |
| 27332 | 50 | 0.0373 | 01 | 18969 | n.a. | n.a. | 0.9627 |
| 27332 | 50 | 0.0373 | 02 | 53121 | 0.1288 | 1.44E+05 | 0.0915 |
| 27332 | 50 | 0.0373 | 03 | 65882 | 0.1062 | 1.88E+03 | 0.0689 |
| 27332 | 50 | 0.0373 | 04 | 94284 | 0.1423 | 6.33E+04 | 0.1050 |
| 27332 | 50 | 0.0373 | 05 | 36875 | 0.0835 | 9.58E+02 | 0.0462 |
| 27332 | 50 | 0.0373 | 06 | 87058 | 0.1229 | 2.20E+04 | 0.0856 |
| 36648 | 50 | 0.0555 | 01 | 22009 | 0.0644 | 1.55E+04 | 0.0089 |
| 36648 | 50 | 0.0555 | 02 | 63177 | 0.0804 | 1.42E+05 | 0.0249 |
| 36648 | 50 | 0.0555 | 03 | 56841 | 0.1203 | 7.48E+03 | 0.0648 |
| 36648 | 50 | 0.0555 | 04 | 82574 | 0.0909 | 1.12E+05 | 0.0354 |
| 36648 | 50 | 0.0555 | 06 | 76484 | 0.0758 | 1.98E+04 | 0.0203 |
| 37398 | 50 | 0.0524 | 01 | 4215 | n.a. | n.a. | 0.9476 |
| 37398 | 50 | 0.0524 | 02 | 12650 | n.a. | n.a. | 0.9476 |
| 37398 | 50 | 0.0524 | 03 | 50598 | 0.0548 | 2.01E+03 | 0.0024 |
| 37398 | 50 | 0.0524 | 04 | 74129 | 0.0926 | 1.17E+05 | 0.0402 |
| 37398 | 50 | 0.0524 | 05 | 1956 | 0.0841 | 1.31E+03 | 0.0317 |
| 37398 | 50 | 0.0524 | 06 | 28046 | 0.1079 | 1.95E+04 | 0.0555 |
| 81804 | 50 | 0.0327 | 01 | 16092 | 0.1294 | 1.89E+04 | 0.0967 |
| 81804 | 50 | 0.0327 | 02 | 13431 | n.a. | n.a. | 0.9673 |
| 81804 | 50 | 0.0327 | 03 | 67782 | 0.1338 | 2.62E+03 | 0.1011 |
| 81804 | 50 | 0.0327 | 04 | 13124 | 0.0964 | 8.09E+04 | 0.0637 |
| 81804 | 50 | 0.0327 | 05 | 61746 | 0.0506 | 1.58E+03 | 0.0179 |
| 81804 | 50 | 0.0327 | 06 | 47310 | 0.0545 | 2.75E+04 | 0.0218 |
| 2913 | 100 | 0.0549 | 01 | 39251 | 0.2534 | 7.30E+03 | 0.1985 |
| 2913 | 100 | 0.0549 | 03 | 39807 | 0.2621 | 1.11E+03 | 0.2072 |
| 2913 | 100 | 0.0549 | 04 | 56532 | 0.2054 | 8.91E+04 | 0.1505 |
| 2913 | 100 | 0.0549 | 05 | 72804 | n.a. | n.a. | 0.9451 |
| 2913 | 100 | 0.0549 | 06 | 39595 | n.a. | n.a. | 0.9451 |
| 50569 | 100 | 0.0482 | 01 | 57899 | 0.1812 | 9.10E+03 | 0.1330 |
| 50569 | 100 | 0.0482 | 02 | 77273 | 0.1500 | 2.12E+05 | 0.1018 |
| 50569 | 100 | 0.0482 | 03 | 35846 | 0.1871 | 1.36E+03 | 0.1389 |
| 50569 | 100 | 0.0482 | 04 | 72475 | 0.1414 | 6.15E+04 | 0.0932 |
| 50569 | 100 | 0.0482 | 05 | 52010 | 0.1669 | 9.59E+02 | 0.1187 |
| 50569 | 100 | 0.0482 | 06 | 60085 | 0.0917 | 1.15E+04 | 0.0435 |
| 74505 | 100 | 0.0393 | 01 | 74396 | n.a. | n.a. | 0.9607 |
| 74505 | 100 | 0.0393 | 02 | 69644 | 0.1002 | 1.14E+05 | 0.0609 |
| 74505 | 100 | 0.0393 | 03 | 79614 | n.a. | n.a. | 0.9607 |
| 74505 | 100 | 0.0393 | 04 | 83242 | 0.1229 | 5.37E+04 | 0.0836 |
| 74505 | 100 | 0.0393 | 05 | 81467 | 0.2571 | 1.09E+03 | 0.2178 |
| 74505 | 100 | 0.0393 | 06 | 62583 | 0.1532 | 1.86E+04 | 0.1139 |
| 78782 | 100 | 0.0511 | 01 | 65698 | 0.1730 | 7.66E+03 | 0.1219 |
| 78782 | 100 | 0.0511 | 02 | 12534 | 0.1120 | 1.02E+05 | 0.0609 |
| 78782 | 100 | 0.0511 | 03 | 6525 | 0.2225 | 1.17E+03 | 0.1714 |
| 78782 | 100 | 0.0511 | 04 | 45156 | 0.1503 | 5.59E+04 | 0.0992 |
| 78782 | 100 | 0.0511 | 05 | 26325 | 0.1778 | 8.72E+02 | 0.1267 |
| 78782 | 100 | 0.0511 | 06 | 51883 | 0.1955 | 1.84E+04 | 0.1444 |
| 91338 | 100 | 0.0602 | 01 | 62676 | n.a. | n.a. | 0.9398 |
| 91338 | 100 | 0.0602 | 02 | 13016 | 0.1190 | 1.24E+05 | 0.0588 |
| 91338 | 100 | 0.0602 | 03 | 62727 | 0.1121 | 9.81E+02 | 0.0519 |
| 91338 | 100 | 0.0602 | 04 | 97899 | 0.1433 | 5.86E+04 | 0.0831 |
| 91338 | 100 | 0.0602 | 05 | 48393 | 0.1808 | 9.40E+02 | 0.1206 |
| 91338 | 100 | 0.0602 | 06 | 6759 | 0.0828 | 7.24E+03 | 0.0226 |